\AtBeginDocument{%
  }

    \documentclass[sigconf]{acmart}

\copyrightyear{2024}
\acmYear{2024}
\setcopyright{acmlicensed}
\acmConference[CSCW Companion '24] {Companion of the 2024 Computer-Supported Cooperative Work and Social Computing}{November 9--13, 2024}{San Jose, Costa Rica.}
\acmBooktitle{Companion of the 2024 Computer-Supported Cooperative Work and Social Computing (CSCW Companion '24), November 9--13, 2024, San Jose, Costa Rica}
\acmISBN{979-8-4007-1114-5/24/11}
\acmDOI{10.1145/XXXXXX.XXXXXX}

\begin{document}

\title{Working with Color: How Color Quantization Can Aid Researchers of Problematic Information}


\author{Nina Lutz}
\authornote{Both authors contributed equally to this research.}
\email{ninalutz@uw.edu}
\affiliation{%
  \institution{University of Washington}
  \city{Seattle}
  \state{Washington}
  \country{USA}
}

\author{Jordyn W. Padzensky}
\authornotemark[1]
\email{jzensky@uw.edu}
\affiliation{%
  \institution{University of Washington}
  \city{Seattle}
  \state{Washington}
  \country{USA}
}

\author{Joseph S. Schafer}
\email{schaferj@uw.edu}
\affiliation{%
  \institution{University of Washington}
  \city{Seattle}
  \state{Washington}
  \country{USA}}

\renewcommand{\shortauthors}{Lutz, Padzensky, and Schafer}

\begin{abstract}
Analyzing large sets of visual media remains a challenging task, particularly in mixed-method studies dealing with problematic information and human subjects. Using AI tools in such analyses risks reifying and exacerbating biases, as well as untenable computational and cost limitations. As such, we turn to adopting geometric computer graphics and vision methods towards analyzing a large set of images from a problematic information campaign, in conjunction with human-in-the-loop qualitative analysis. We illustrate an effective case of this approach with the implementation of color quantization towards analyzing online hate image at the US-Mexico border, along with a historicist trace of the history of color quantization and skin tone scales, to inform our usage and reclamation of these methodologies from their racist origins. To that end, we scaffold motivations and the need for more researchers to consider the advantages and risks of reclaiming such methodologies in their own work, situated in our case study. 
\end{abstract}



\received{23 May 2024}

\maketitle

\section{Introduction}
Scholars of problematic information, such as misinformation and online hate campaigns, often rely on large corpuses of social media data. Many of these datasets have not been largely visual, although we know that visual media has a key role in such information campaigns. At scale, analysis of visual media sets requires new methodologies to be brought into CSCW and social computing spaces. Artificial intelligence (AI) and machine learning (ML) methods hold great potential to aid in such analyses, yet have presented biases and analysis gaps, particularly towards marginalized groups \cite{buolamwini_gender_2018}. In the context of misinformation and online hate, using these methods can become particularly problematized, as we do not want to reify or exacerbate underlying biases, nor miss out on key data points for how marginalized individuals are being targeted by hateful and misleading rhetoric. Additionally, many of these technologies are costly and computationally prohibitive. 

As such, we look towards lighter weight, non-AI and non-ML methods from computer vision and graphics, to enable better exploration and analysis of image datasets in problematic information campaigns. Such methods vary, and include approaches such as intraset cosine similarity to compare how images in datasets deviate from one another quantitatively, as used in the study of text-to-image generator outputs in auditing their bias \cite{ghosh-caliskan-2023-person} to averaging images together to understand biases in large datasets \cite{buolamwini_gender_2018}. 

We extend these approaches in our ongoing, mixed-methods, and interpretivist case study of visual media regarding the US-Mexico border crisis. To analyze and debunk this misinformation, we utilize these quantitative image analysis methods with human-in-the-loop qualitative coding cooperative workflows to understand the role of visual media in this particular problematic information campaign. 

This particular contribution focuses on one method from this pipeline: color quantization, a method by which summative color palettes and color information can be derived from a digital image. We examine the history of this non-AI and non-ML method as applied to human subjects, and how we may incorporate it into our pipeline for analyzing datasets and better sampling of images for human qualitative coding. To this end, we ask two research questions:
\begin{enumerate}
    \item RQ1: What insights can be gained from non-machine learning methodologies to the study of visual misinformation about the U.S.-Mexico Border? 
    \item RQ2: How can we carefully and thoughtfully reclaim these methods for critical and empirical research at scale of crises events involving people? 
\end{enumerate}

To answer these questions, we contribute the following: 
\begin{enumerate}
\item A critical, historical informed framework of using color quantization in mixed-methods, human-in-the-loop pipelines
\item A use case of our color quantization implementation with empirical results from our case study 
\end{enumerate}

In doing so, we position this work as a critical inquiry, reclaiming a non-ML image analysis method for large-scale image collections. In the wake of the AI summer \cite{bruckman_surviving_2024}, we push back and say that we do not believe these tools are fully equipped to handle the problems that image-based misinformation presents, and we instead take a human-in-the-loop approach with thoughtful implementation and integration of quantitative methods.

\section{Related Work}

\subsection{Visual Misinformation} \label{vismis}
The misinformation field has traditionally relied heavily on textual analysis of social media posts \cite{oh_community_2013, bliss_agenda_2020}. But given the increase of visual-first social networks and technologies of AI-generated imagery, misinformation research – particularly research that follows interpretive traditions to understand the role of misinformation in large social and political contexts \cite{oh_anxiety_2013, prochaska_mobilizing_2023, kharazian_our_2024} – stands to benefit significantly from incorporating visual analysis at scale into this work to address some of these platforms’ novel challenges \cite{wang_culturally-embedded_2019}.

Misinformation is not a new area of study, building off decades of research \cite{jack_lexicon_nodate, kharazian_our_2024, shibutani_improvised_1966} of studying large rumoring campaigns, now largely coined mis- and disinformation, with misinformation as the umbrella terminology of such misleading, propagating information narratives. Such studies of these large-scale online rumors combine methodologies from various fields such as communications, political science, sociology, and computer science. These mixed-methods, interpretivist traditions have been utilized to study large amounts of misinformation that can do true harm on and offline \cite{muhammed_t_disaster_2022}, often through the study of problematic information which may contain hate, violence, and explicit content to fuel these narratives and conspiracy theories \cite{cinelli_dynamics_2021, kim_misinformation_2021}. 

In CSCW and related spaces that study online misinformation, common techniques involve building graphs of networks spreading information to understand the interactions between bad actors and unwitting spreaders \cite{starinfluence, beers_followback_2023} and utilizing qualitative coding and content analysis to draw out campaign themes \cite{starbird_could_2016, prochaska_mobilizing_2023}. This occurs through a process of sensemaking deeply rooted in organizational studies \cite{weick}, where audiences, but also researchers, make sense of campaigns of information \cite{dailey_its_2015, Mamykina}. 

As humans cannot possibly code all posts related to a campaign, quantitative techniques such as natural language processing \cite{oshikawa_survey_2020}, network analyses \cite{beers_followback_2023}, and engagement data visualization \cite{jasser_dynamics_2019} are applied to get larger-scale trends about these narratives and aid with exploratory data analysis in this mixed-methods work \cite{prochaska_mobilizing_2023}. There has been work in visual methods for such exploratory data analysis of image collections, such as the work of Lev Manovich, who draws insights and new representations from large collections of images \cite{manovich_chapter_2020}.  Additionally, there is increasing interest in sensemaking and qualitative coding in visual media as related to crises, political events, and misinformation  \cite{7192668, medina_serrano_dancing_2020}. However, there is a need for further work in this space. 

Along with how visual media contributes to misinformation campaigns and trends within these collections, there is also a spectrum of “cheap fakes” to “deep fakes” \cite{paris_deepfakes_2019} and the need to investigate human and AI manipulation and creation of media. The field of media forensics utilizes quantitative, computational methodologies to understand the tampering or synthesizing of digital visual media (photographs, videos, GIFs, etc) \cite{farid_image_2020, bhagtani_overview_2022, bohme_multimedia_2009}, doing just this. There is ample opportunity to bring more computational methods into misinformation campaign studies to perform baseline media forensics, such as leveraging color quantization to see if an image has been edited by color space inconsistencies. We demonstrate this case later in this work. 

\subsection{Historicism in CSCW and Color Quantization}
CSCW focuses on technologies and systems of cooperative and collaborative work that pre-date modern computers. Sensemaking and rumoring have long histories before internet mis- and disinformation, and the idea color schemas and relationships to human skin long predate digital mixing and color spaces like RGB. We ground our work in historicism within CSCW, an area that has been identified as under-utilized \cite{soden_time_2021}, to understand how looking back at not just theories and underpinnings but uses of technology and methods can influence how we study and view phenomena today – and add a sensibility towards countering what we may see as present, progress, and novelty. 

For many scholars in CSCW space, skin tone and imagery have come to a major head within the last decade, with advances such as scholars Joy Buolamwini and Timnit Gebru’s pivotal “Gender Shades” paper exposing how computer vision algorithms incorporated into high stakes technical systems often ignore darker skin tones \cite{buolamwini_gender_2018}. This utilized the Fitzpatrick scale to measure these discrepancies, but acknowledged its limitation but that it was the best and most reasonable measure available. In subsequent years, various studies have examined the biases against and problematization of skin tone categories and representations \cite{heldreth_which_2024, thong2023skin}. 

However, such skin tone scales have a lineage tracing back to 18th-century anthropology, where Johann Blumenbach is often credited with being amongst the first to formally and academically introduce phenotypical skin tone categories \cite{bhopal_beautiful_2007, rupke_johann_2018}. Blumenbach posited 5 racial groups and went on to become the father of scientific racism, proclaiming an evolutionary superiority of white people based on cranial geometry and color theory \cite{bhopal_beautiful_2007, rupke_johann_2018}. This paved the way for the work of Von Luschan, who in the early 20th century developed the Hautfarbe Tafel \cite{swiatoniowski_comparing_2013} a tool for systematically identifying 36 skin tones by region, based on their deviation from whiteness, and went on to be used by the Nazis in WWII eugenics movements \cite{holocaust_museum}. 

Long after scientific racism was less accepted, there was still a drive for categorization and summarization of human skin tones for medical and scientific purposes. The Fitzpatrick scale directly adopted the Von Luschan scale to make a scale of human skin tones with respect to skin cancer risks in 1975 \cite{goon2021skin}. This was later found to be flawed, leading to the under diagnosis of cancer in people with darker skin tones \cite{goon2021skin}. Regardless, scale has become ubiquitous, becoming the standard of emojis across many platforms \cite{miltner_one_2021}. Many CSCW and social computing scholars have studied its role in self-presentation and digital identity via emojis \cite{robertson_self-representation_2018, robertson_black_2021, wang_culturally-embedded_2019} and data visualization scholars have considered its repercussions on how human data is represented outside of emojis \cite{dhawka_we_2023}. 

In the last few years, there have been advancements toward improving the representation and analysis of human skin tone shades, both within and outside of academic pursuits. Fenty’s 2017 makeup launch with a record number of 40 foundation shades, particularly servicing a variety of customers with darker skin, increased and has forever reset the standard for diversity of skin tone coverage in makeup lines \cite{werle2019beyond}. Building off the flaws of the Fitzpatrick scale and answering the calls of bias against darker skin tones in machine learning and camera systems, the Monk Skin Tone Scale, in collaboration with Google, was developed as a new and improved standard of skintone annotation for machine systems, informed by social science and historical methods \cite{monk_monk_2023}.

Such quantified skin tone classification schemas will never fully capture the range of human skin, but methodologies of quantification to begin such measurements may benefit investigations such as ours in identifying the potential presence of human subjects across large datasets, particularly towards capturing a larger variety of skin tones in such datasets to expose and understand the narratives against people of color.

\subsection{Digital Color Quantization}
Color quantization is the digital process by which distinct colors of image pixels are quantified into color spaces by representing them in 3D geometric vectors \cite{orchard_color_1991, brun2017color}. As such, this process can find averages and trends in the colors of an image, and reduce the colors of an image for image compression and analysis. 

This process occurs via projecting these vectors into geometric color spaces, such as RGB (red, green, and blue). The RGB space, derived from the trichromatic theory of human color perception, represents all colors as numerical functions of red, green, and blue  and has become a technological standard in many digital and physical displays \cite{trichrome, susstrunk1999standard}. However, this space can obscure the visibility of key colors, such as representing browns and pinks as mathematically similar, although they are dissimilar to human perception. Color spaces such as HSV utilize units of hue, saturation, and value to solve this, making all hues (i.e. yellow vs blue) as geometrically having their own planes, making analysis of pixel distributions easier at first glance, and more intuitive to human vision, optimizing its use for image editing and analysis \cite{shuhua2010application}. 

\begin{figure}[h]
  \centering
  \includegraphics[width=\linewidth]{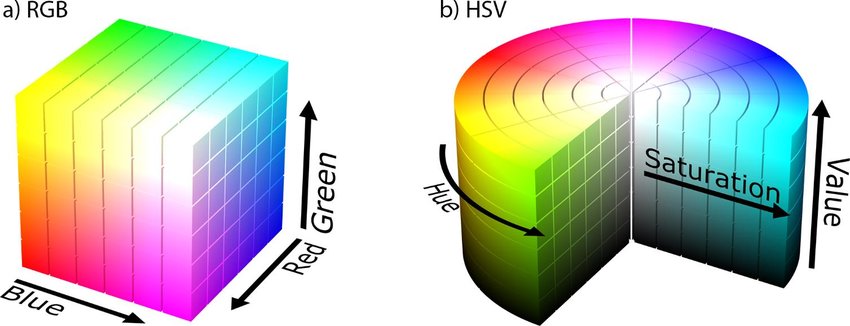}
  \caption{Showing color spaces in RGB and HSV geometries. Figure credit: Michael Horvath \cite{wiki}.}
  \Description{Figure showing an RGB cube color space and HSV cylinder color space.}
\end{figure}

\begin{figure}[h]
  \centering
  \includegraphics[width=\linewidth]{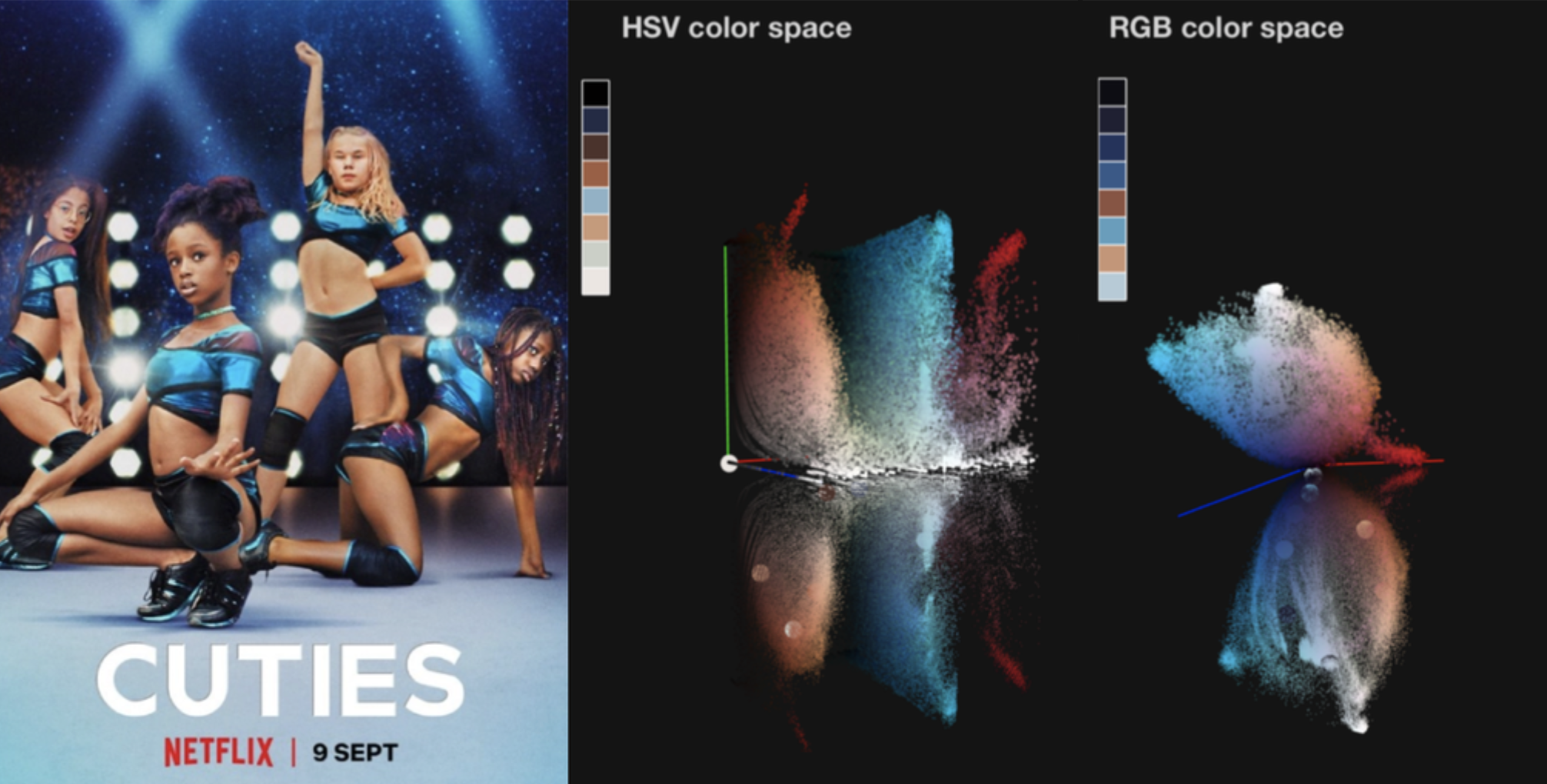}
  \caption{Differences between HSV and RGB spaces in image color quantization. While the summative palette remains the same, the left HSV space distinguishes the hues for more intuitive human summarization. Source; Lutz, 2020 \cite{lutz_cuties_2020}. }
  \Description{Figure showing the Cuties movie poster in color quantization distributions in HSV and RGB spaces.}
\end{figure}

In digital images, each pixel has an encoded color. In many programming libraries, like OpenCV, this color can be easily extracted in either RGB or HSV values \cite{opencv_library}. Alternatively, these values can be converted between one another mathematically in code, by projecting RGB vectors to HSV space and vice versa \cite{chernov2015integer}. In order to implement color quantization, these values are stored in one data structure – quantizing them from abstract pixels to an ordered collection of vectors. Analysis may be performed on this set, commonly toward understanding a smaller representative subset of colors to describe this image. This can be done in many ways, such as finding global maximums in color histograms or utilizing k-means clustering and reporting out the cluster centroids \cite{orchard_color_1991}.

\section{Methods}
\subsection{Color Quantization Implementation}
We have implemented a color quantization protocol in Python to take in image(s) to aid analysis in our case study. This code offers three functionalities: 
\begin{enumerate}
  \item Constructs the color quantization image distributions and summative color palette with k-means clustering 
  \item Flags images if it is suspected that humans are within them based on the Monk skin tone scale  
  \item If a particular object or symbol of interest is likely to exist in the image
\end{enumerate}

\begin{figure}[h]
  \centering
  \includegraphics[width=\linewidth]{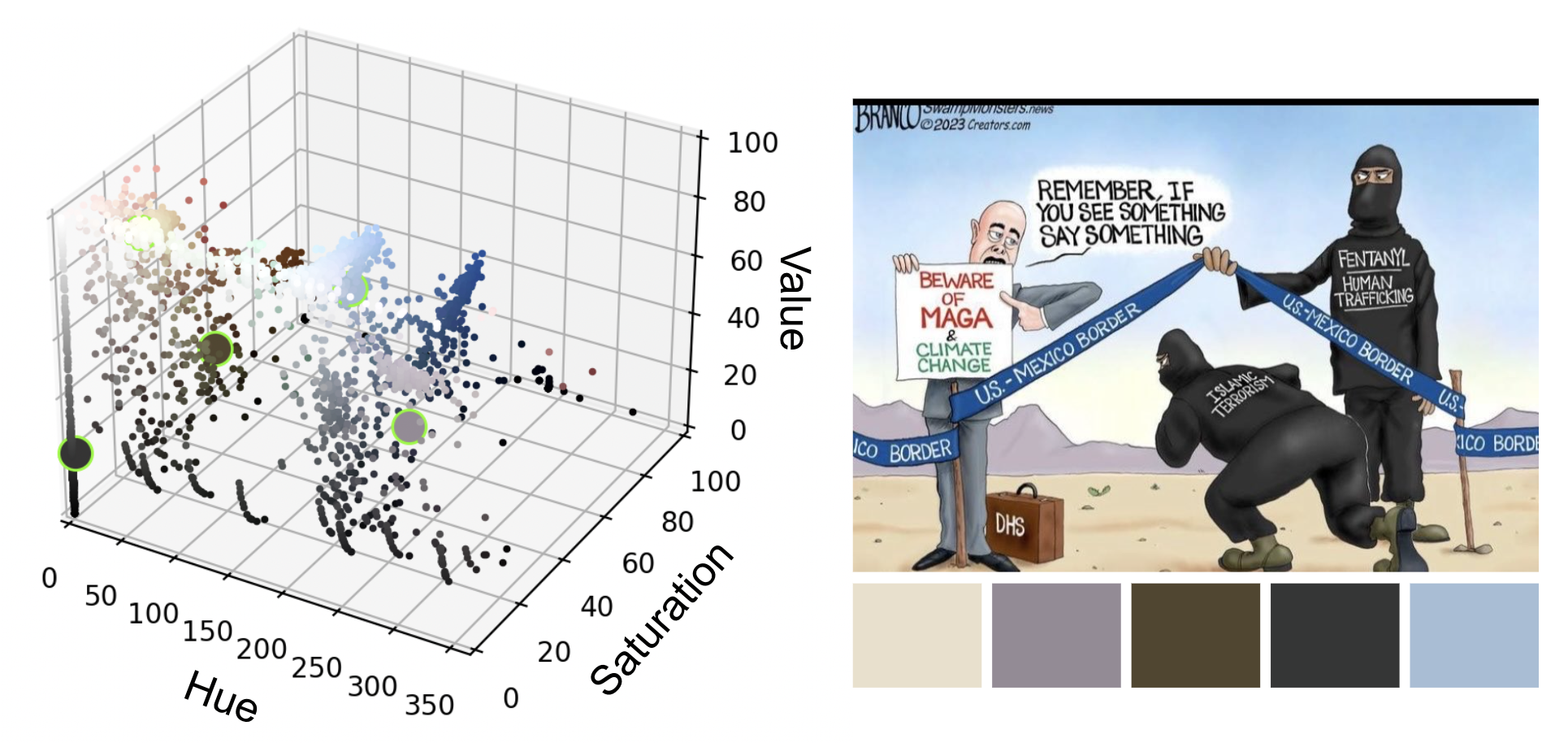}
  \caption{ On the left, the pixel color distribution of the image on the right in HSV color space, with centroids plotted as larger points. The summative color palette for the image is below the image itself. This image was selected from our sample set of imagery on misinformation and online hate about the US-Mexico border rhetoric.}
  \Description{On the left, the pixel color distribution of the image on the right in HSV color space, with centroids plotted as larger points. The summative color palette for the image is below the image itself.}
  \label{fig:example}
\end{figure}

This functionality supports the analysis of large-scale image collections in projects like ours, which have humans in the loop and seek thematic analysis and descriptive statistics regarding the image collection for further analysis. The overall color quantization is a fairly standard implementation \cite{orchard_color_1991, brun2017color}. It takes the following steps, the results of which are captured in Figure \ref{fig:example}: 
\begin{enumerate}
\item Image is uploaded in Python with the OpenCV library \cite{opencv_library}.
\item Converts each pixel to HSV color space, storing them in an ordered list, based on the pixel ordering from the default OpenCV structure \cite{opencv_library}.
\item K-means clustering, with a user identified k\footnote{We recommend k in the range of 4-6 based on our trials and palette results.}, is then performed to cluster the colors, with the k centroids of these clusters saved to another list.
\item A distribution of pixels are then plotted in 3D space with the centroids visually denoted for quick debugging. 
\end{enumerate}

It is worth noting that the centroids themselves can also be returned alone, representing a summative color palette. Additionally, this code can either be batched or done on an image of choice for media forensics settings, explained in \ref{mediafor}.
 
\subsection{Human Subject or Chosen Object Implementation}
To signify if a human or other object of interest may be in the image, we can search for pixel recurrences and ratios of certain colors, through a correlational search of particular color ratios in relation to one another \cite{huang_spatial_1999, keating1975improved, smith1996tools}. This allows us to search extensive collections of images with a low weight, rapid query for ones that we suspect contain either human subjects or objects of interest. This method is lighter weight and removes potential biases by not relying on existing recognition approaches. However, this methodology will still have errors, which are acceptable in our implementation given its goal of being leveraged to aid our human qualitative coders to explore image collections. 

To detect humans, we utilize the Monk Skin Tone scale color, bounding it by 15\% on each dimensional side of the color in HSV space to adjust for shadows and lightning differences and capture the orb distribution from their scale for annotation \cite{monk_monk_2023}. These are meant to serve as tonalities, not exact skin shades as the Fenty 40 point scale would be \cite{heldreth_which_2024}, and thus more appropriate for more lighting conditions for our particular study, as per Monk et al.’s design of this scale as a guardrail and bucketing \cite{heldreth_which_2024, monk_monk_2023}. Such approximations can help flag images that potentially incorporate human subjects for review. Given that our qualitative coding or analysis does not look for skin tones, these are not units of measure but rather a methodology to capture as many potential human subjects as possible.

\begin{figure}[h]
  \centering
  \includegraphics[width=\linewidth]{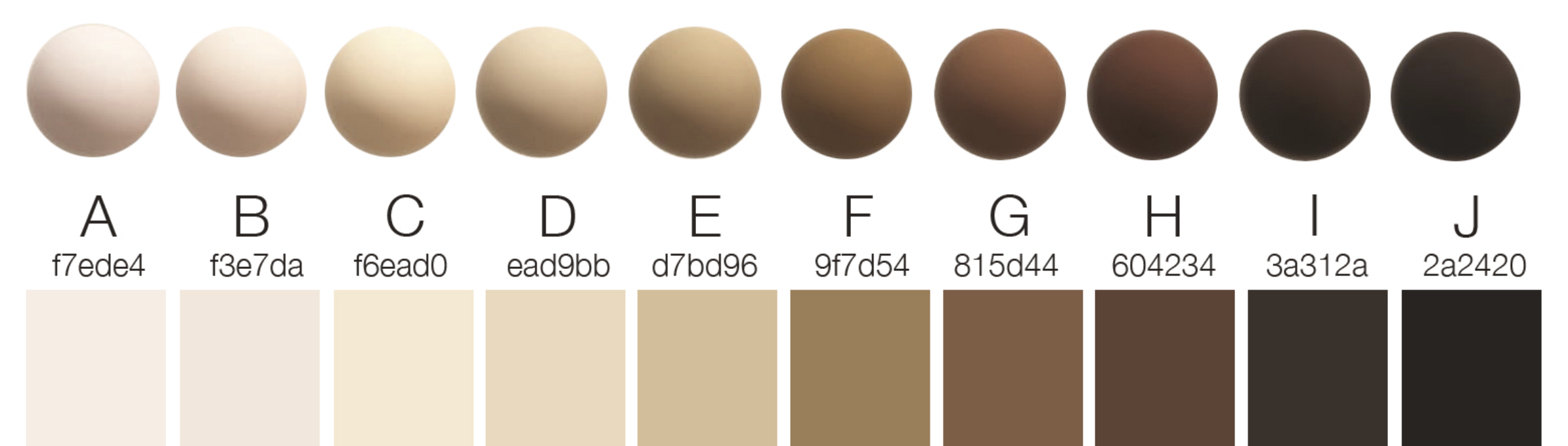}
  \caption{Monk scale and orb palettes. We use  starting HEX colors and projected them into HSV value ranges with 15\% bound on all axes permutations, to capture colors across the orb distribution and within our images. Figure credit: Dr. Ellis Monk \cite{monk_monk_2023}.}
  \Description{Figure showing 3D orbs and 2D rectangular color swatches of the Monk scale.}
  \label{fig:monk}
\end{figure}

For the detection of objects of key interest, we focus on symbols of hate groups for our particular research, which we sourced from the Global Extremist Symbols Database maintained by the nonprofit Global Project Against Hate and Extremism \cite{noauthor_hate_nodate}. For any given image or symbol, either from the database or another object of interest, we run color quantization on it, soliciting summative palettes to search for within an image, as seen in Figure \ref{fig:PB}. We then utilize the same search approach as human subjects to flag if an image may contain an object of interest based on clusters of pixels being similar to the palette. We suspect that this approach could be used for other kinds of key objects as well, but focus on hate symbols in our case study.

\subsection{Mixed-Methods Human-in-the-Loop Content Analysis}
This implementation of color quantization for summarization and exploration is meant for integration into a human-in-the-loop pipeline. In cases of problematic information campaigns, humans often qualitatively code samples of media as part of content analysis \cite{mayring2015qualitative}. In our current study of harmful rhetoric at the US-Mexico border, a codebook has been inductively developed from the trending media we collected in this space. We utilize Label Studio \cite{Label}, an open-source software similar to AtlasTI, to code our images with this codebook. 

This color quantization allows us to better sample from our overall dataset, to get a more representative sample for human qualitative coders to identify the most salient points in this narrative based on our research questions, prioritizing how migrants are depicted in these narratives at the US-Mexico border. We can also sample for hate groups or particular objects of interest such that qualitative coders can explore if an image is misappropriated or has additional context. 

Along with the qualitative coding samples, this method also allows us to analyze collections of images that are too large for coders to code, by outputting descriptive statistics and summary trends for the image collection. 

\section{Results}
Our research team has had success with this lightweight and flexible methodology towards supporting the collaborative work of qualitative coding of US-Mexico border rhetoric. It has allowed our qualitative coders to gather more relevant and representative samples of images for analysis, gather descriptive statistics and summaries, and incorporate media forensics into our work. 

\subsection{Human and Select Subjects of Interest}
Color quantization, like any method, is not foolproof, and will have margins of error in over or under-selecting images. Our current case study found an error rate of approximately 18.4\% for human subjects\footnote{For our purposes, human subjects do not include far range crowd photos where individuals are not discernible, a common image that was excluded from this measurement.
} in photographic images, with a tendency to over-select images (false positives) and an error rate of 9.6\% for relevant hate symbols in illustrative and photographic images. We arrive at this error bound from a test run on 2000 images in our dataset for 5 hate symbols. 

\begin{figure}[h]
  \centering
  \includegraphics[width=\linewidth]{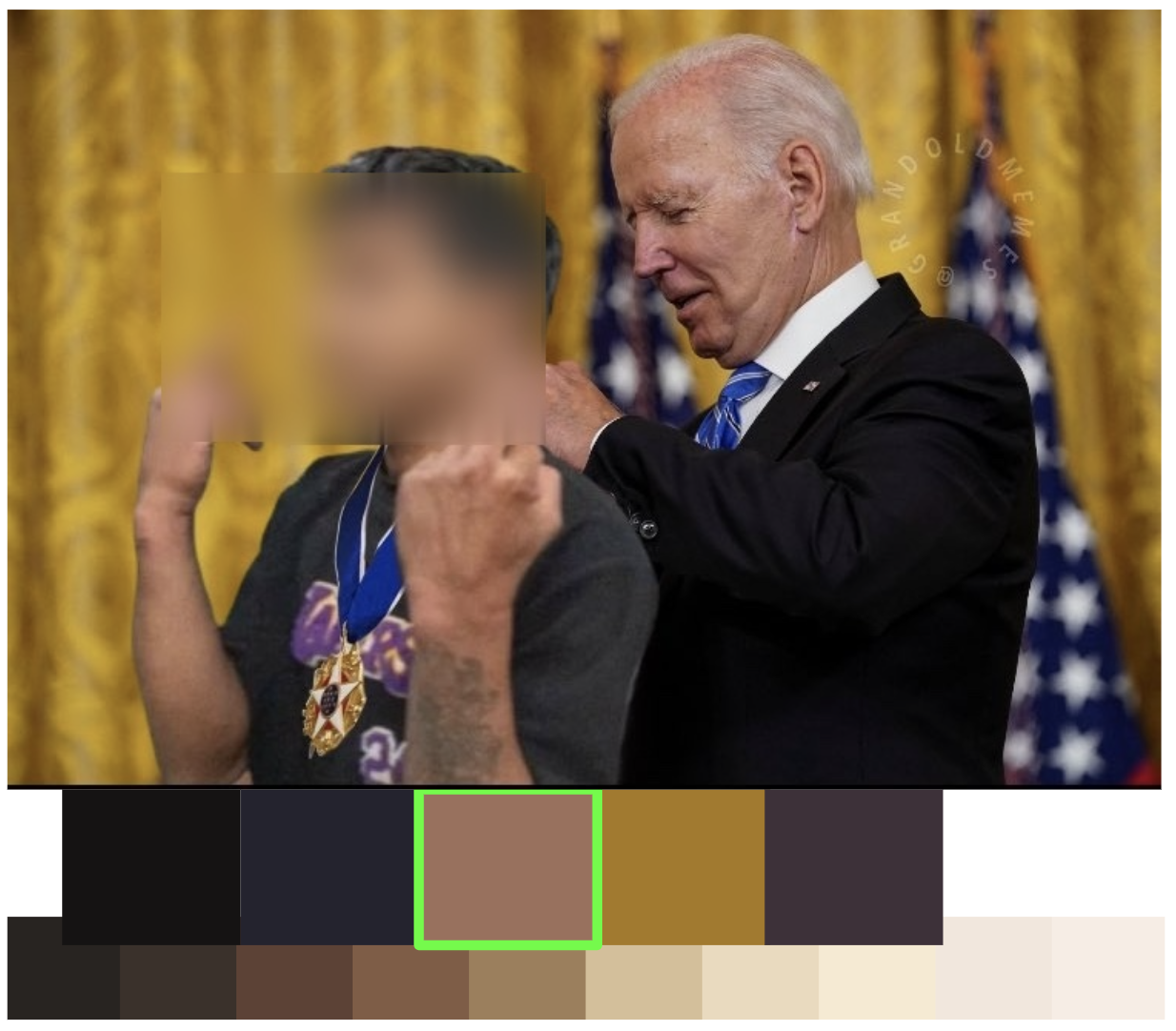}
  \caption{Summative palette extracted from Biden image, with the green highlighted color falling into the Monk Skin Tone Spectrum (below palette). Face blurred for privacy \cite{schafer_screenshot_2024}. This image was Photoshopped and is not an accurate image, and has been used as part of an online hate campaign against Latin American migrants.}
  \Description{Figure of a photoshopped, fake image of Biden giving a migrant a medal. Below is the derived color palette with a highlighted square showing the color matched to the Monk Skin Tone Scale.}
  \label{fig:biden}
\end{figure}

\begin{figure}[h]
  \centering
  \includegraphics[width=\linewidth]{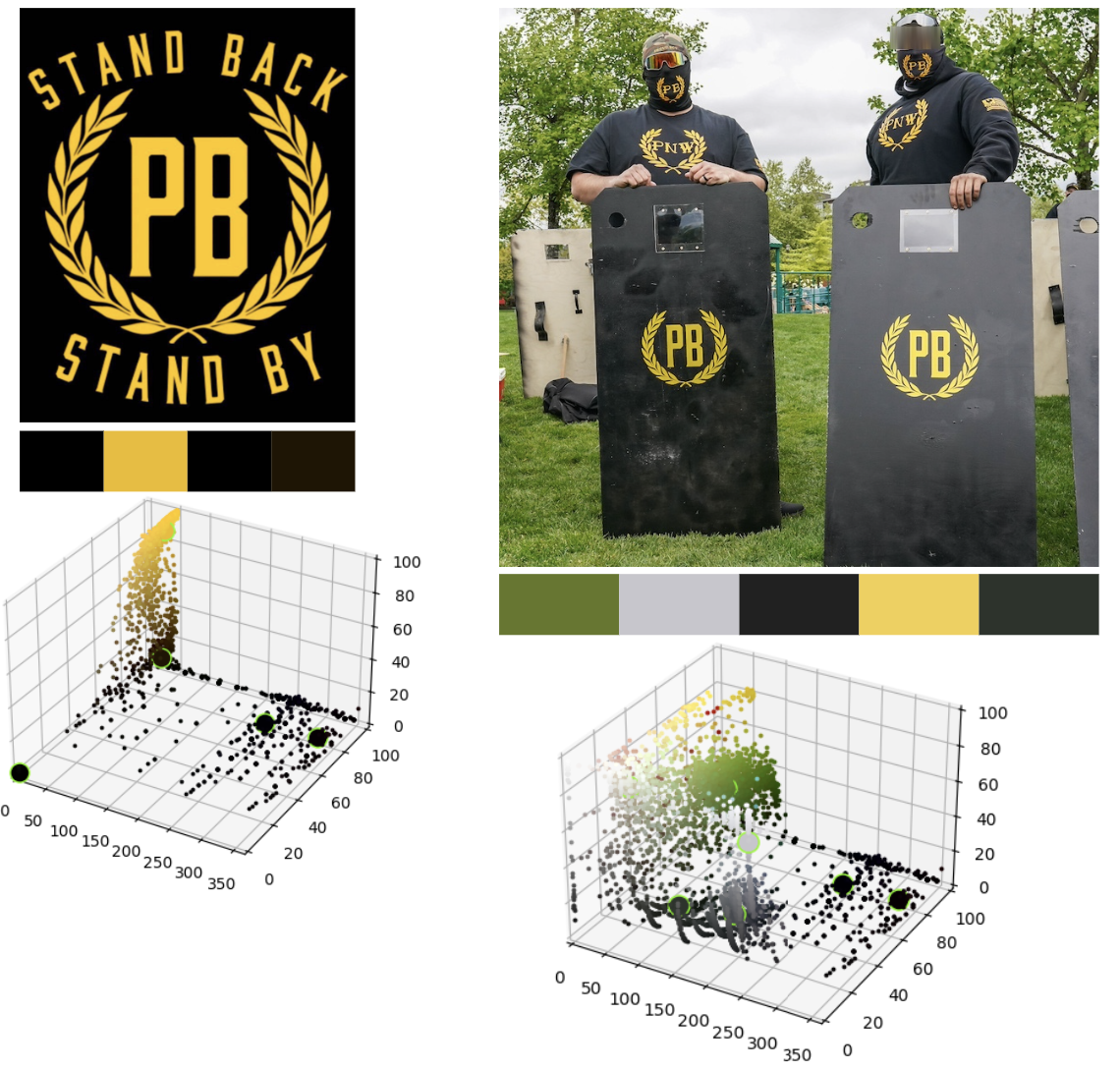}
  \caption{Image on right flagged for hate imagery due to clusters of yellow pixels around proportionately more black and black adjacent pixels, as per the color distribution from the logo taken from the Global Extremist Symbols Database \cite{noauthor_hate_nodate}. Eyes blurred for privacy \cite{schafer_screenshot_2024}.}
  \Description{Figure showing a logo of the Proud Boys, a US hate group, being detected in an image of two men in these logos. Below both images are the color palettes and the color distributions.}
  \label{fig:PB}
\end{figure}

Regardless, we have found this methodology helpful in a larger human-in-the-loop pipeline to study these large-scale campaigns towards identifying human subjects in images, as seen in Figure \ref{fig:biden}, or objects of interest, such as Figure \ref{fig:PB}. Such images are flagged by the code for their pixel relationships to either the Monk skin tones or an object of interest from our database.

This method is not robust enough for a fully automated analysis. In our pipeline, to ensure nuance is added to these quantifiers, human coders verify and contextualize these images. 

\subsection{Human-in-the-loop Interaction}
This code is lightweight and efficient with very few dependencies, making it easy to integrate into a larger pipeline. This setup is particularly useful in aiding a human-in-the-loop approach, with methods such as exploratory data analysis, descriptive statistics, and improving qualitative coding. 
Our team uses this mixed-methods pipeline, where exploratory data analysis is paramount. Using this method allows us to attain high-level summaries, such as the frequency by which certain subjects may appear, and allows us to leverage color quantization data to cluster images into visually similar groupings for future study and sampling. All this aids the qualitative coding experience by giving human coders a better bird's-eye view of their data than what exploratory text methods such as word clouds may otherwise give. 

\subsection{Media Forensics}\label{mediafor}
\begin{figure}[h]
  \centering
  \includegraphics[width=\linewidth]{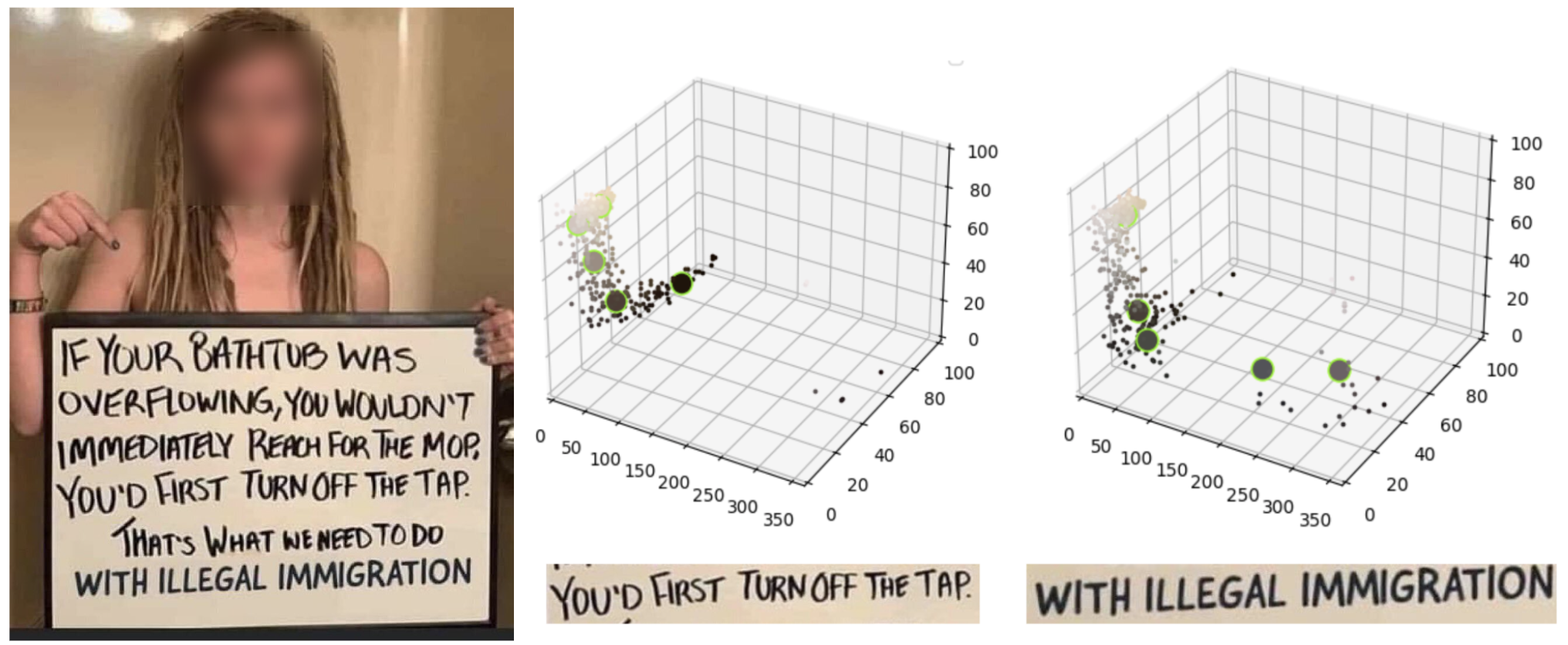}
  \caption{Different color spaces of the lines in this image show quantitative evidence that the last line could not be physically from the photo, supporting the qualitative observation. Face blurred for privacy \cite{schafer_screenshot_2024}}
  \Description{A picture of a woman holding a sign that is photoshopped, with color quantization showing that last line has a different color distribution than another line.}
  \label{fig:mediafor}
\end{figure}

There are times in such human-in-the loop projects where team members may need to prove if an image was altered, such as to explicitly call out misleading and harmful actors. There are a plethora of media forensics methods for detecting image manipulation, aforementioned in \ref{vismis}. Color quantization can also be a media forensics tool, by seeing if regions of an image match the color distribution to see if they are from the same image or if the image has been tampered with. Figure \ref{fig:mediafor} features an image that qualitative coders marked as tampered. However, analysis of the color quantization and color channels show that the last line of text was added to this image and is not from this same image, given that its color space is drastically different than the other lines. This means this line of text could not have been physically made with the same marker, lighting, and surface, supporting the qualitative observation. 

\section{Discussion}
There is an increased and important need for critical theory and methods regarding the treatment of people of color in extensive and complex sociotechnical systems powered by AI and emerging computational technologies \cite{ogbonnaya-ogburu_critical_2020}. Many of these models and methods reify biases and exacerbate harms in working and environmental conditions \cite{anthony2020carbontracker, gray2019ghost, klaus, buolamwini_gender_2018}. Thus, in our investigation, we argue for CSCW researchers seeking methods to parse large collections of data, particularly of human subjects, to reflect and apply criticality and historicism \cite{soden_time_2021} and ask if they need to use blackbox tools and models over lighter-weight, physics-informed approaches such as color quantization. We posit such methods as a beneficial alternative in research regarding sensitive and marginalized populations, such as problematic information and hate content. However, these solutions are not without their own flaws, and studies using these lighter-weight techniques must still proceed with rigorous care for studied populations.

While these visual methods may not rely on current issues around AI models and their climate costs, they are derived from dark histories which have still become ubiquitous, such as adoption in  emojis \cite{miltner_one_2021}.  What does it mean to reclaim and discuss methodologies used to historically harm and marginalize people of color? What does it mean to reclaim them when in the backdrop, they have become so baked into technology and visual culture? When we use this method to study xenophobic misinformation at the US-Mexico border, we are truly utilizing a point of a lineage that once aided, justified, and made scientific racism de facto and de jour, and allowed the colonization and socioeconomic disenfranchisement of individuals that are now migrating largely from the Global South to the Global North. Therefore, reclaiming the method of color quantization is not simple. As Audre Lorde cautions against utilizing methods of domination to dissemble it, which is stated in her oft-cited warning “The master's tools will never dismantle the master's house.” \cite{lorde2018master}. 

Such challenges remain open for further scholarship and development, alongside the assumption that there is a need to perform color quantization and summation of human skin tones themselves. However, we hope this application and intellectual contribution will pave forward a path to challenge CSCW scholars to step back and interrogate alternatives to AI-enabled analysis of visual images, along with the complexities of these sociotechnical alternatives. 

Along with the criticality of such approaches to datasets, there comes a point where the value of humans in the loop becomes paramount. For decades, qualitative, human coding has been used in conjunction with computational methods to bring rich insights around how people interact at scale in digital spaces \cite{bordia_problem_2004, vieweg_collective_2008, starbird_could_2016}. In our study, this approach allows images to be flagged for qualitative coders as potential images with human subjects or objects of interest. This allows for better sampling but also for better analysis, where researchers may be able to consider new questions in such datasets. Furthermore, in cases of extreme hate content, this methodology may assist in protecting qualitative researchers from the worst of the worst in their data \cite{kelley_researching_2020, schafer_participatory_2023}. 

\section{Limitations}
Color quantization, and by extension geometric pixel matching, is a convenient method for these types of tasks in many ways – it runs fast and it is fully dependent on geometry and color spaces as opposed to models of object detection or AI tools. However, for this particular use case, this method could benefit from some select model implementations, such as technologies like MediaPipe \cite{mediapipe} which can identify humans with greater accuracy or other object recognition approaches \cite{sukanya2016survey, objdeeplearning, surveyobj}. Color quantization is additionally limited in identifying human subjects where no skin is visible, where other geometric methods may thrive. There are also limitations in objects of interest, such as “alternate color” logos, which may have different color spaces. There are other object detection methods that can be implemented with machine learning or other graphics methodologies which may serve better for identifying objects in such cases where our system may struggle. Regardless, this method provides an extremely durable, lightweight, and useful way for researchers to rapidly explore large amounts of images in studying problematic information campaigns. 

\section{Conclusion and Future Work}
We have presented our implementation of color quantization for use in a mixed-methods human-in-the-loop pipeline to study the visual rhetoric of problematic information campaigns, using empirical evidence and test results from our case study of the visual media surrounding the U.S.-Mexico Border on social media. We have demonstrated the functionality of this methodology and how it enhances our collaborative workflow within our research team for insights and analysis regarding a large visual dataset. Furthermore, we have shown the critical and theoretical merit of incorporating historic visual methods, which have typically been previously used to invoke harm or have been radically exclusionary. 

Towards our own work, we will continue iterating and enhancing such methodologies into our pipelines, by incorporating methods with color quantization but also beyond it. Going forward, we hope that more researchers not only explore visual information campaigns online in mixed methods ways, but also consider the histories and careful implementations and limits of their methods. It is our sincere ambition that such inquiries will develop more rich visual research methodologies and theories at the intersection of CSCW and visual culture. 

\section{Acknowledgements}
We are grateful to Professor Kate Starbird for supporting this work. We are also grateful to our colleagues involved in this larger research project from whom we received feedback: Katie Arriaga, Mert Bayar, Wilson Chen, Dominic Monteperto, Logan Tuttle, and Tiffany Yan. Nina Lutz has been supported in conducting this by an NSF SaTC Grant (No. 2120496). Joseph S. Schafer has been supported by an NSF Graduate Research Fellowship (No. DGE-2140004). Any opinions, findings, conclusions, or recommendations expressed in this material are those of the authors and do not necessarily reflect the views of the National Science Foundation or other funders.

\bibliographystyle{ACM-Reference-Format}
\bibliography{1_bib}



\end{document}